\documentclass[twocolumn,showpacs,showkeys,preprintnumbers,prd,nofootinbib,superscriptaddress]{revtex4-1}

\usepackage{amsmath}
\usepackage{amsfonts}
\usepackage{amssymb}
\usepackage{graphicx}
\usepackage{color}
\usepackage[]{hyperref}
\usepackage{booktabs}
\usepackage{cleveref}

\Crefname{equation}{Eq.}{Eqs.}
\Crefname{figure}{Fig.}{Figs.}
\Crefname{section}{Sec.}{Secs.}

\begin{document}

\title{The phase-space view of non-local  gravity cosmology}

\author{Salvatore Capozziello}
\email{capozziello@na.infn.it}
\affiliation{Dipartimento di Fisica ``E. Pancini", Universit\`a di Napoli ``Federico II", Via Cinthia 9, 80126 Napoli, Italy.}
\affiliation{Scuola Superiore Meridionale, Largo S. Marcellino 10, 80138 Napoli, Italy.}
\affiliation{Istituto Nazionale di Fisica Nucleare (INFN), Sez. di Napoli, Via Cinthia 9, 80126 Napoli, Italy.}

\author{Rocco D'Agostino}
\email{rocco.dagostino@unina.it}
\affiliation{Scuola Superiore Meridionale, Largo S. Marcellino 10, 80138 Napoli, Italy.}
\affiliation{Istituto Nazionale di Fisica Nucleare (INFN), Sez. di Napoli, Via Cinthia 9, 80126 Napoli, Italy.}

\author{Orlando Luongo}
\email{orlando.luongo@unicam.it}
\affiliation{Dipartimento di Matematica, Universit\`a di Pisa, Largo B. Pontecorvo, 56127 Pisa, Italy.}
\affiliation{Universit\`a di Camerino, Divisione di Fisica, Via Madonna delle carceri, 62032 Camerino, Italy.}
\affiliation{NNLOT, Al-Farabi Kazakh National University, Al-Farabi av. 71, 050040 Almaty, Kazakhstan.}

\begin{abstract}
We consider non-local Integral Kernel Theories of Gravity in a homogeneous and isotropic universe background as a possible scenario to drive the cosmic history. In particular, we investigate the cosmological properties of a gravitational action containing the inverse d'Alembert operator of the Ricci scalar proposed to improve Einstein's gravity at both high and low-energy regimes. In particular,  the dynamics of a physically motivated non-local exponential coupling is analyzed in detail by recasting the cosmological equations as an autonomous system of first-order differential equations with dimensionless variables.
Consequently, we study the phase-space domain and its  critical points, investigating  their stability and main properties. In particular, saddle points and late-time cosmological attractors  are discussed in terms of the free parameters of the model. Finally, we discuss the main physical consequences of our approach in view of dark energy behavior and the $\Lambda$CDM model.
\end{abstract}  

 
\maketitle

\section{Introduction}

The missing ingredients in our understanding of the cosmological puzzle,  namely dark matter \cite{unoo} and dark energy \cite{dueo}, may suggest the idea of developing theories of gravity beyond
General Relativity (GR) \cite{Faraoni}. Even though the success of GR is confirmed by a wide number of experiments, the question of whether GR could be considered the ultimate theory for describing the gravitational interaction is raised by several shortcomings at both infrared and ultraviolet scales.

At the infrared scales, the simplest way to take into account the late-time acceleration of the universe suggested by current observations \cite{Riess98,Perlmutter99,Copeland06,Haridasu17,Planck} is to introduce the cosmological constant $(\Lambda)$ into the Einstein-Hilbert action, mimicking the effects of an anomalous fluid with negative pressure, dubbed  dark energy. This represents the main constituent of the universe energy budget and it  is at the base of the cosmological concordance model, namely the $\Lambda$CDM model. However,
the $\Lambda$CDM model is severely plagued by theoretical and conceptual issues \cite{Weinberg89,Sahni02,Peebles03,ROM}, such as the \emph{fine-tuning} problem arising from the very high discrepancy between the observed value of $\Lambda$ and the vacuum energy density predicted by Quantum Field Theory (QFT), and the \emph{coincidence} problem related to the question of why the magnitudes of $\Lambda$ and matter densities are so close today.

If from the one hand, dark energy has been investigated through different approaches, including dynamical scalar fields \cite{Ratra88,Caldwell98,Armendariz00}, unified dark energy models \cite{Peebles99,Scherrer04,Liddle08,Anton-Schmidt,Brandenberger21,D'Agostino22} and holographic dark energy scenarios \cite{Li04,Tavayef18,D'Agostino19,Saridakis20}, on the other hand, the attempt to unify GR and Quantum Mechanics results in non-renormalizable ultraviolet divergences at the two-loop level \cite{Goroff86}. Moreover, QFT and GR do not easily reconcile since gravity cannot be unified under the same standard of other fundamental interactions, so a full quantum theory of gravity is far from being achieved.

To overcome the infrared and ultraviolet issues, Extended Theories of Gravity (ETGs) have been extensively proposed throughout the last two decades  \cite{Capozziello11,Nojiri17,review}. Such theories include the presence of geometric invariants (other than the standard Ricci curvature scalar, linear in the action) and/or scalar fields non-minimally coupled to geometry. These can be seen as   semi-classical approaches where effective actions emerge to model out interactions related, for example,  to quantum fields in curved spaces \cite{Birrell}. GR is recovered in some limit or as a particular case in wide classes of theories. Particular focus has been devoted to theories such as $f(R)$, $f(T)$ and $f(Q)$ to reproduce the dark energy features \cite{Capozziello:2002rd,DeFelice10,Nojiri11,rocco_f(R),Bengochea09,Linder10,rocco_f(T),Beltran18,Bajardi20,Frusciante21,Anagnostopoulos21,rocco_f(Q), Cai}. In these cases, besides the standard  curvature representation of dynamics, related to the Ricci scalar $R$, it is possible to consider some equivalent representations as the teleparallel formulation of gravity, related to the torsion scalar $T$, and the non-metric formulation, based on the non-metricity scalar $Q$. See \cite{Heisenberg} for details.

A recent, intriguing  possibility, in the framework of ETGs, consists in breaking the locality principle. In fact, the effective actions of all fundamental interactions show a dynamical non-locality at the one-loop level, and a  description of Quantum Gravity  may be provided by non-local theories of gravity \cite{Bajardi1,Bajardi2,Modesto,Briscese13}. A  family of non-local theories of gravity is represented by the so-called  Infinite Derivatives Theories of Gravity (IDGs), which consider analytic functions of a differential operator, such as the covariant d'Alembert operator, into the gravitational action.  It has been shown that IDGs  are able to heal Big Bang and black hole singularities \cite{Modesto,Briscese13}. Another  family is the Integral Kernel Theories of Gravity (IKGs) that  take into account   integral kernels of differential operators, such as the inverse d'Alembert operator \cite{Deser07,Maggiore14,Nesseris14,Belgacem18}. IKGs are inspired by infrared corrections in QFT \cite{Barvinsky15} and  emerge in searching for  unitary and  renormalizability of quantum gravity models \cite{Biswas12, Biswas:2005qr, Biswas:2016etb, Biswas:2016egy,Buoninfante:2018xiw,Buoninfante:2018mre}.
After the seminal paper \cite{Deser07}, non-local gravity acquired a lot of interest in cosmology being considered a natural mechanism to address dark energy dynamics and, eventually, dark matter issues in the large-scale structure \cite{Filippo}.

In this paper, we are going to consider  cosmological realizations of IKGs where the  Lagrangian density is given by the general function $f(R,\Box^{-1}R)$ \cite{Bajardi2}.  Such models represent natural non-local extensions of $f(R)$ gravity,  considered  to explain current acceleration and inflation at once. The aim is to  account \emph{de facto} for both infrared and ultraviolet quantum corrections \cite{Nojiri08,Koivisto08}. In particular, we  are going to study the phase-space portrait of this scenario to investigate the dynamics in terms of critical points and cosmological stability at late times. Finally, we discuss the possible physical consequences  of a non-local  paradigm for background cosmology.

The structure of the paper is as follows.  In \Cref{sec:grav},  we briefly sketch  non-local gravity effective action giving the related field equations.  \Cref{sec:cosm} is devoted to  cosmological dynamics related to non-local gravity. Specifically,  setting the cosmological background as a spatially-flat Friedman-Lema\^itre-Robertson-Walker (FLRW) universe, we derive the related autonomous dynamical system to develop the phase-space view and find out the  critical points (\Cref{sec:critical}). The stability analysis  in view to derive the  late-times cosmic attractors is presented in \Cref{stability}. Discussion and conclusions are reported in \Cref{sec:conclusions}. Throughout this work, we use physical units such that $c=1=\hbar$.

\section{Non-local gravity in a nutshell}
\label{sec:grav}

As discussed above, a straightforward approach to deal with non-locality in gravity is introducing non-local corrections in the gravitational action. The simplest proposal is considering the Hilbert-Einstein action, linear in the Ricci scalar $R$, plus terms where non-local operators are  present.
We focus on the following non-local gravity action \cite{Nojiri08,Koivisto08}:
\begin{align}
S =  \int d^4 x \sqrt{-g}\left\{\dfrac{R}{2\kappa}\left[1+f\left(\Box^{-1}R\right)\right] +\mathcal{L}_m \right\},
\label{eq:action_nl}
\end{align}
where $\kappa\equiv 8\pi G$, $g$ is the determinant of the metric tensor $g_{\mu\nu}$ and $\Box \equiv \nabla^\mu\nabla_\mu$ is the d'Alembert operator. Here, $f$ is an arbitrary function, while $\mathcal{L}_m$ represents a matter field Lagrangian density.

Introducing two generic scalar fields $\phi$ e $\xi$, it is possible to obtain the localized form of action \eqref{eq:action_nl}:
\begin{equation}
S  =  \int d^4 x \sqrt{-g}\left\{\dfrac{1}{2\kappa}\left[R(1+f(\phi)-\xi) - \nabla_\alpha\xi\nabla^\alpha\phi\right] + \mathcal{L}_m\right\}\,.
\label{eq:action_localized}
\end{equation}
In the Appendix, we discuss the conditions for the above theory to be ghost-free \cite{Nojiri:2010pw}. 

Varying \Cref{eq:action_localized} with respect to $\xi$ one finds
\begin{equation}
 \Box\phi-R=0  \Longleftrightarrow \phi=\Box^{-1}R\,.
 \label{eq_motion_1}
\end{equation}
In other words, the auxiliary field behaves like Lagrange multipliers.

For the sake of convenience, let us  introduce a new field $\psi \equiv f(\phi) - \xi$ and recast the action under the form
\begin{equation}
S =  \int d^4 x \sqrt{-g}\bigg\{\frac{1}{2\kappa}\left[R(1+\psi) - f_\phi(\nabla\phi)^2 +\nabla_\mu\psi\nabla^\mu\phi\right]+\mathcal{L}_m \bigg\},
\label{eq:action}
\end{equation}
where $f_\phi \equiv \partial f/\partial \phi$. The above expression for the action represents a two-scalar-tensor theory with a scalar field, $\psi$, nonminimally coupled to gravity.

The field equations are obtained by varying Eq.~\eqref{eq:action} with respect to the metric tensor $g_{\mu\nu}$, giving
\begin{align}
&R_{\mu\nu}(1+\psi)-\dfrac{1}{2}g_{\mu\nu}\left[R(1+\psi)-f_\phi(\nabla\phi)^2+\nabla_\sigma\psi\nabla^\sigma\phi-2\Box\psi\right] \nonumber \\
&+\nabla_\mu\psi\nabla_\nu\phi-\nabla_\mu\nabla_\nu\psi-f_\phi \nabla_\mu\phi\nabla_\nu\phi= \kappa T_{\mu\nu}\,,
\end{align}
while varying the action \eqref{eq:action} with respect to $\phi$ yields
\begin{equation}
\Box\psi-f_{\phi\phi}(\nabla\phi)^2-2Rf_\phi=0\,,
\end{equation}
where the last term on the left-hand side is obtained by using Eq.~\eqref{eq_motion_1}.

As discussed in \cite{Bajardi1,Bajardi2}, the form of the function $f(\phi)$ can be fixed requiring the presence of Noether symmetries in dynamics. 
In what follows, we shall study the cosmological solutions with non-local corrections and investigate the stability of the corresponding dynamical behavior.

\section{Cosmological dynamics}
\label{sec:cosm}

Under the hypothesis of a homogeneous and isotropic universe, we consider the  FLRW metric, $ds^2=-dt^2+a(t)^2\delta_{ij}dx^idx^j$, with zero spatial curvature to fulfill current observations. Here  $a(t)$ is the cosmic scale factor, depending upon the cosmic time only. Consequently, the scalar fields $\psi$ and $\phi$ also depend on time only, and their dynamics obey the following equations of motions, respectively:
\begin{subequations}
\begin{align}
\ddot{\psi}+ 3H\dot{\psi}& =f_{\phi\phi}\dot{\phi}^2-12f_\phi(\dot{H} + 2H^2)\,,  \label{eq:psi} \\
\ddot{\phi}+ 3H\dot{\phi}& =-6(\dot{H} + 2H^2)\,. \label{eq:phi}
\end{align}
\end{subequations}
Thus, the Friedmann equations are 
\begin{align} 
3H^2&= \dfrac{1}{1 + \psi}\left[\kappa \rho_m + \dfrac{1}{2}(f_\phi \dot{\phi}^2 - \dot{\psi}\dot{\phi})-3H\dot{\psi}\right], \label{eq:first Friedmann} \\
-2\dot{H} - 3H^2&= \dfrac{1}{1 + \psi}\left[\kappa p_m + \dfrac{1}{2}(f_\phi \dot{\phi}^2 - \dot{\psi} \dot{\phi})+ \ddot{\psi} + 2H\dot{\psi}\right], \label{eq:second Friedmann}
\end{align}
where  $\rho_m$ and $p_m$ are the density and pressure of matter, respectively which obey the continuity equation
\begin{equation}
\dot{\rho}_m+3H(1+w_m)\rho_m=0\,,
\label{eq:continuity}
\end{equation}
being $w_m\equiv p_m/\rho_m$ the equation of state (EoS)  parameter of matter.

The cosmological dynamics  can be efficiently studied through an autonomous system of first-order differential equations \cite{Capozziello_Phase, Bahamonde_review}. In order to do that, we define the dimensionless variables
\begin{subequations}
\begin{align}
&x\equiv \dfrac{\dot{\phi}}{6H}\,, \hspace{1cm} y\equiv -\dfrac{\dot{\psi}}{H(1+\psi)}\,,\\
& z\equiv \dfrac{6f_\phi}{1+\psi}\,, \quad \Omega_m\equiv \dfrac{\kappa \rho_m}{3H^2(1+\psi)}\,.
\end{align}
\end{subequations}
We note that the matter density parameter $\Omega_m$ is not an independent variable, since it can be determined once the values of $(x,y,z)$ are known. Specifically, from \Cref{eq:first Friedmann}, one finds the constraint
\begin{equation}
1=\Omega_m+ x^2z + xy + y\,.
\label{eq:constraint}
\end{equation}
Therefore, combining \Cref{eq:psi,eq:phi,eq:first Friedmann,eq:second Friedmann,eq:continuity}, after some algebra, one obtains the following dynamical system in terms of the new variables:
\begin{subequations}
\begin{align}
& x'= \frac{1}{2} \big[w_\text{eff}(1+x)-3x-1\big], \\
& y'= y^2+\frac{3}{2} (w_\text{eff}-1) y+z\Big(1-\frac{6f_{\phi\phi}x^2}{f_\phi}-3 w_\text{eff}\Big), \\ 
& z'= z \left(\frac{6 f_{\phi\phi} x}{f_\phi}+y\right), \\
& \Omega_m'=\Omega_m \big[y+3 (w_\text{eff}-w_m)\big].
\end{align}
\end{subequations}
Here, the prime denotes derivative with respect to the number of e-folds $N\equiv\ln a$ and $w_\text{eff}$ is the effective EoS parameter, given by
\begin{align}
w_\text{eff}&\equiv-1-\dfrac{2\dot{H}}{3H^2} \nonumber \\
& =-1+\frac{1}{3 (1-z)}\bigg[\frac{6 f_{\phi\phi} x^2 z}{f_\phi}+3 (w_m+1) \Omega_m \nonumber \\
&\hspace{3cm}+6 x^2 z+6 x y+4 y-4 z\bigg].
\end{align} 
It is worth  stressing that acceptable solutions are those for $z\neq1$.

We now consider the universe filled with nonrelativistic pressureless matter, so that $w_m=0$. Moreover, we assume a coupling of the form 
\begin{equation}
f(\phi)=f_0\,e^{\alpha\phi}\,,
\label{ansatz}
\end{equation}
where $f_0$ and $\alpha$ are constants. It can be shown that a non-local  exponential coupling  naturally emerges requiring the existence of Noether symmetries into dynamics \cite{Bajardi1,Bajardi2,Bahamonde17}.  It is worth noticing that this kind of coupling assumes a particular role in the super-renormalizability of Quantum Gravity  as discussed in  \cite{Modesto}. In other words, the non-local exponential coupling can suitably represent the IR behavior of the gravitational field starting from  UV scales.

In this case, $f_{\phi\phi}/f_\phi=\alpha$  and the dynamical system reads 
\begin{subequations}
\begin{align}
& x'= \frac{1}{2} \big[w_\text{eff}(1+x)-3x-1\big], \label{eq:x}\\
& y'= y^2+\frac{3}{2} (w_\text{eff}-1) y+z\left(1-6\alpha x^2-3 w_\text{eff}\right),   \label{eq:y}\\ 
& z'= z \left(6\alpha x+y\right),   \label{eq:z}\\
& \Omega_m'=\Omega_m \left(y+3w_\text{eff}\right),  \label{eq:Omega}
\end{align}
\end{subequations}
where
\begin{equation}
w_\text{eff}=-1+\frac{4 (y-z)+6 x\left[y+xz(1+\alpha)\right]+3 \Omega_m}{3(1-z)}\,.
\end{equation}
In the following, we investigate the critical points of the above system in order to search for possible attractor solutions.

\section{The phase-space view}
\label{sec:critical}

\begin{table*}
\begin{center}
\setlength{\tabcolsep}{0.4em}
\renewcommand{\arraystretch}{2}
\begin{tabular}{c c c c c c c c c}
\hline
\hline
Critical point & $(x, \, y, \, z)$ &   $\Omega_m$ & $w_\text{eff}$ & Existence\\
\hline
I  & $\left(-\frac{1}{3},\, 0, \, 0\right)$  & 1  & 0 & $\forall \alpha$   \\
II & $\left(\frac{1}{2} \big(-1+\sqrt{1-\frac{2}{3 \alpha}}\big), \,3 \alpha -\sqrt{3\alpha(3\alpha-2)}, \,0	\right)$ & 0 & $1-2 \alpha +2 \sqrt{\alpha ^2-\frac{2 \alpha }{3}}$ & $\alpha<0 \vee \alpha\geq \frac{2}{3}$ \\
III & $\left(-\frac{1}{2} \big(1+\sqrt{1-\frac{2}{3 \alpha}}\big), \,3 \alpha +\sqrt{3\alpha(3\alpha-2)}, \,0	\right)$  & 0 & $1-2 \alpha -2 \sqrt{\alpha ^2-\frac{2 \alpha }{3}}$ & $\alpha<0 \vee \alpha\geq \frac{2}{3}$\\
IV & $\Big(-\frac{1}{3 \alpha},\, 2, \, 3\alpha \left(2-3\alpha\right)\Big)$ & 0 & $\frac{\alpha -1}{3 \alpha -1}$  &  $\alpha\neq 0 \vee \alpha\neq \frac{1}{3}$ \\
V &  $\left(\frac{2 \alpha }{6 \alpha ^2-3 \alpha -\eta },\, \frac{12 \alpha ^2}{3 \alpha -6 \alpha ^2+\eta },\, \frac{9 \alpha -6 \alpha ^2-\eta}{2}\right) $ & $\frac{2 \alpha  \left[\alpha  (78 \alpha ^2-39 \alpha+9-11 \eta)+3 \eta \right]}{\left(3 \alpha -6 \alpha ^2+\eta \right)^2}$ & $\frac{1}{6}\left(3-6 \alpha-\frac{\eta}{\alpha}\right)$ & $\alpha\neq 0$\\
\hline
\hline
\end{tabular}
\caption{Critical points with corresponding matter density and effective EoS parameter for the $f(\phi)=f_0 e^{\alpha\phi}$ model.  Here, $\eta=\sqrt{3 \alpha ^2 \left(3-4 \alpha +12 \alpha ^2\right)}$. See \Cref{sec:critical} for the discussion.}
 \label{tab:critical}
\end{center}
\end{table*}
The phase-space view of a dynamical system mainly consists in detecting the fixed points and analyzing the trajectories around them.

Specifically, the fixed (critical) points of the above autonomous system \eqref{eq:x}--\eqref{eq:Omega} are found from solving $x'=y'=z'=0$. In particular, we shall search for solutions describing a matter-dominated universe $(\Omega_m=1\,, w_\text{eff}=0)$ and accelerated universe $(-1<w_\text{eff}<-1/3)$, including the case of a de Sitter universe $(w_\text{eff}=-1,\,  \Omega_m=0)$.

Hence, let us consider the first fixed point:
\begin{equation}
(x,y,z)_I=\left(-\dfrac{1}{3},\, 0, \, 0\right).
\label{eq:matter-dominated}
\end{equation}
In correspondence to this point, we have $\Omega_m^{(I)}=1$ and $w_\text{eff}^{(I)}=0$, which clearly describes a matter-dominated universe solution.

Then, the second fixed point is 
\begin{equation}
(x,y,z)_{II}=\left(\frac{1}{2} \Big(-1+\sqrt{1-\frac{2}{3 \alpha}}\Big), \,3 \alpha -\sqrt{3\alpha(3\alpha-2)}, \,0	\right),
\end{equation}
for $\alpha<0 \vee \alpha\geq 2/3$. Hence, one finds
\begin{align}
\Omega_m^{(II)}&=0\,, \\
w_\text{eff}^{(II)}&=1-2 \alpha +2 \sqrt{\alpha ^2-\frac{2 \alpha }{3}}\,.
\end{align}
In this case, $\nexists\, \alpha$ such that the second critical point can provide an accelerated universe.

\noindent The third fixed point is 
\begin{equation}
(x,y,z)_{III}=\left(-\frac{1}{2} \Big(1+\sqrt{1-\frac{2}{3 \alpha}}\Big), \,3 \alpha +\sqrt{3\alpha(3\alpha-2)}, \,0	\right),
\end{equation}
for $\alpha<0 \vee \alpha\geq 2/3$. We thus obtain
\begin{align}
\Omega_m^{(III)}&=0\,, \\
w_\text{eff}^{(III)}&=1-2 \alpha -2 \sqrt{\alpha ^2-\frac{2 \alpha }{3}}\,.
\end{align}
In the interval $2/3<\alpha<3/4$, we have accelerated solutions. In particular, $\alpha=3/4$ corresponds to a de Sitter universe, fully-dominated by a pure cosmological constant, namely $w_\text{eff}=-1$, for which the critical point reads
\begin{equation}
(x,y,z)_{III}^{(dS)}=\left(-\frac{2}{3}, \, 3, \, 0 \right).
\end{equation}

\noindent The fourth fixed point is
\begin{equation}
(x,y,z)_{IV}=\left(-\frac{1}{3 \alpha},\, 2, \, 3\alpha \left(2-3\alpha\right)\right),
\end{equation}
for $\alpha\neq 0 \vee \alpha\neq 1/3$. In this case, we have
\begin{align}
\Omega_m^{(IV)}&=0 \,, \\
w_\text{eff}^{(IV)}&=\frac{\alpha -1}{3 \alpha -1}\,.
\label{eq:w_eff}
\end{align}
If $\alpha=1/2$, we find again a de Sitter universe with pure cosmological constant, whose critical point reads 
\begin{equation}
(x,y,z)_{IV}^{(dS)}=\left(-\frac{2}{3},\, 2, \, \frac{3}{4}\right).
\label{eq:de-Sitter}
\end{equation}

\noindent The fifth fixed point is
\begin{equation}
(x,y,z)_{V}=\left(\frac{2 \alpha }{6 \alpha ^2-3 \alpha -\eta },\, \frac{12 \alpha ^2}{3 \alpha -6 \alpha ^2+\eta },\, \frac{9 \alpha -6 \alpha ^2-\eta}{2}\right),
\end{equation}
for $\alpha\neq 0$ and $\eta=\sqrt{3 \alpha ^2 \left(3-4 \alpha +12 \alpha ^2\right)}$. Then, we obtain
\begin{align}
\Omega_m^{(V)}&=\frac{2 \alpha  \left[\alpha  (78 \alpha ^2-39 \alpha+9-11 \eta)+3 \eta \right]}{\left(3 \alpha -6 \alpha ^2+\eta \right)^2}\,, \\
w_\text{eff}^{(V)}&=\frac{1}{6}\left(3-6 \alpha-\frac{\eta}{\alpha}\right).
\end{align}
In this case, accelerated solutions exist for $1/3 \leq \alpha \leq 3/4$. A de Sitter universe is found for $\alpha=3/4$, in correspondence to which the critical point is given as
\begin{equation}
(x,y,z)_{V}^{(dS)}=\left(-\frac{2}{3},\, 3, \, 0\right).
\end{equation}
The sixth fixed point is 
\begin{equation}
(x,y,z)_{VI}=\left(\frac{2 \alpha }{6 \alpha ^2-3 \alpha +\eta },\frac{12 \alpha ^2}{3 \alpha -6 \alpha ^2-\eta }, \frac{9 \alpha -6 \alpha ^2 +\eta}{2} \right),
\end{equation}
for $\alpha\neq 0$. However, it is easy to show that  $\Omega_{m}^{(VI)}>1$, $\forall \alpha\neq0$, implying a nonphysical solution. Thus, the sixth critical point will not be further considered  in our analysis.

\section{The stability analysis }
\label{stability}

\begin{table*}
\begin{center}
\setlength{\tabcolsep}{0.4em}
\renewcommand{\arraystretch}{2}
\begin{tabular}{c c c c c c c c c}
\hline
\hline
Critical point & Eigenvalues &  Stability \\
\hline
I  & $\left(-\frac{3}{2}, \, -\frac{3}{2}, \, -2 \alpha\right)$  & Stable for $\alpha>0$, saddle for $\alpha<0$ \\
II  & $\left(0, \, 0, \, 3-3 \alpha + \sqrt{3\alpha  (3 \alpha -2)}\right)$  & Unstable \\
III  & $\left(0, \, 0, \, 3-3 \alpha - \sqrt{3\alpha  (3 \alpha -2)}\right)$  & Unstable \\
IV  &  $\left(\frac{2-3 \alpha }{1-3 \alpha}, \, \frac{2-3 \alpha }{1-3 \alpha}, \, \frac{5-9 \alpha }{1-3 \alpha }\right)$ & Stable for $\frac{1}{3}<\alpha <\frac{5}{9}$, saddle for $\frac{5}{9}<\alpha <\frac{2}{3}$, unstable for $\alpha<\frac{1}{3} \vee \alpha>\frac{2}{3}$ \\
V & -- & Saddle for $\alpha<0 \vee 0<\alpha<\frac{5}{9} \vee \alpha >\frac{3}{4}$, unstable for $\alpha=\frac{3}{4}$\\
\hline
\hline
\end{tabular}
\caption{Eigenvalues and stability of the critical points for the $f(\phi)=f_0 e^{\alpha\phi}$ model.}
 \label{tab:stability}
\end{center}
\end{table*}

Let us now study the stability of  critical points summarized in \Cref{tab:critical}. In particular, we check whether the  previously obtained cosmological solutions may represent late time attractors. To this end, we shall compute linear perturbations of the dynamical system and analyze the sign of the eigenvalues of the Jacobian matrix once calculated in correspondence of each fixed point. 
We can distinguish among three possible cases: if the real parts of the eigenvalues are all negative, we have a stable (attractor) point; if they are all positive, they correspond to an unstable (repeller) point; if their signs are mixed, the corresponding critical point is a saddle point.

\begin{figure*}
\includegraphics[width=3.2in]{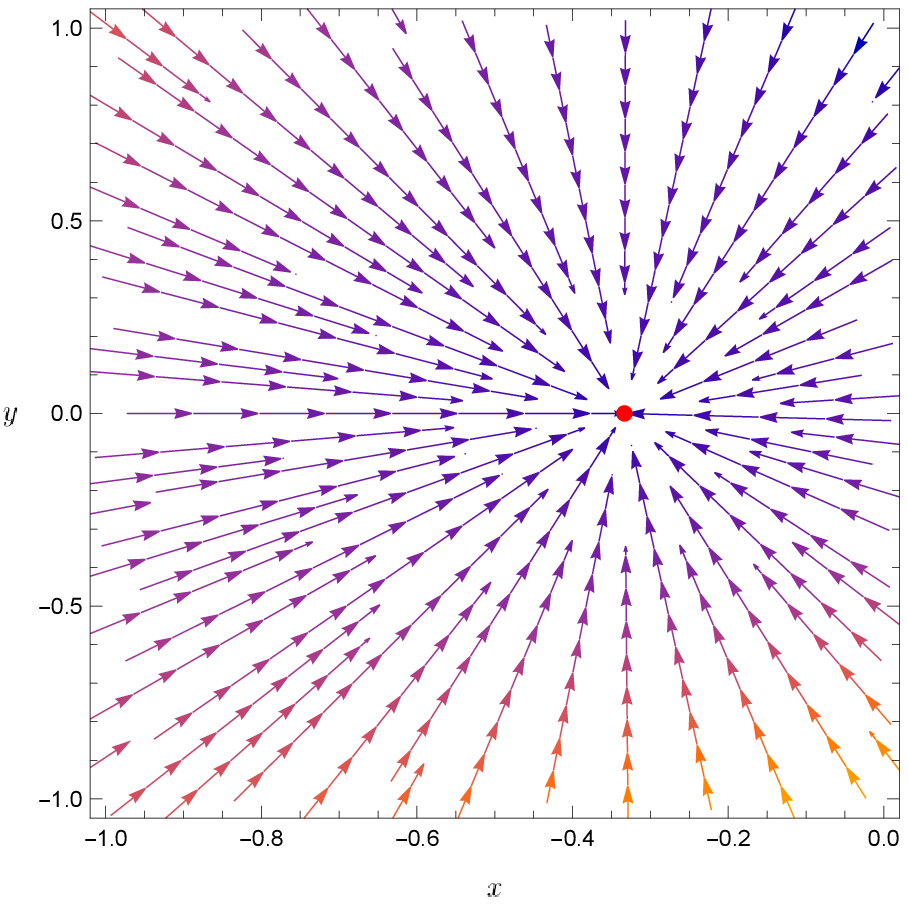}
\hspace{0.7cm}
\includegraphics[width=3.2in]{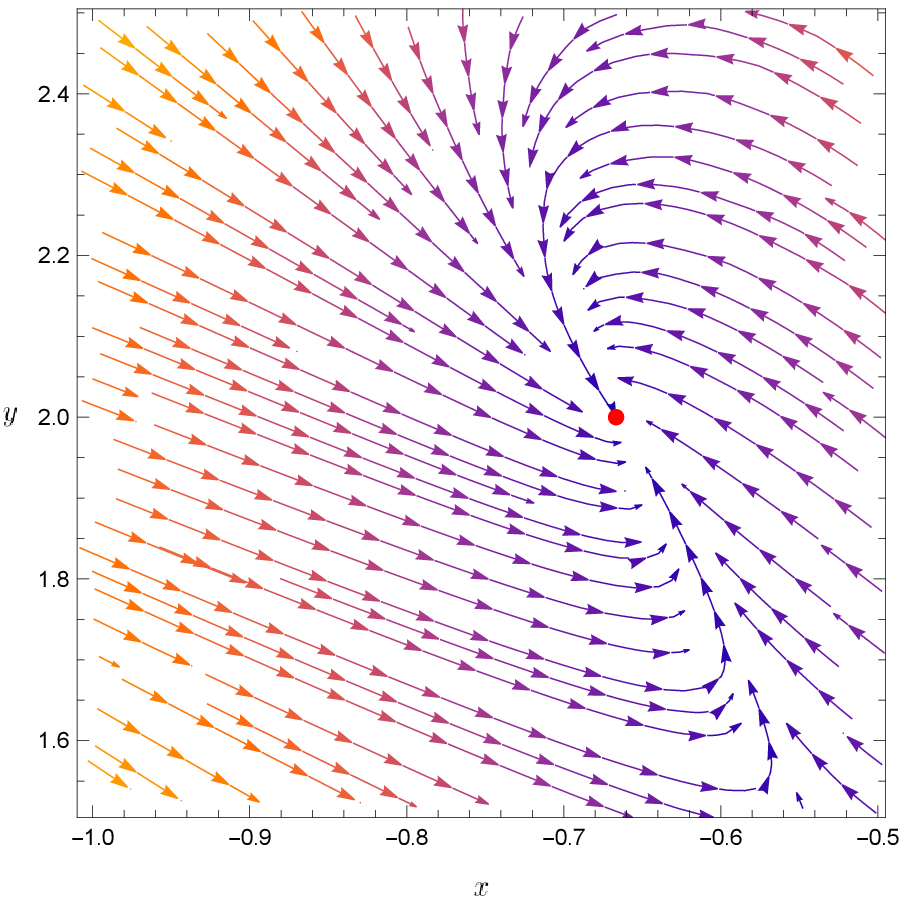}
\caption{Phase-space trajectories around the attractors of the $f(\phi)=f_0 e^{\alpha\phi}$ model. The red dots correspond to the matter-dominated solution \eqref{eq:matter-dominated} emerging from the first critical point (left panel) and the de Sitter solution \eqref{eq:de-Sitter} emerging  from the fourth critical point (right panel).}
\label{fig:attractors}
\end{figure*}

Linear perturbations of \Cref{eq:x,eq:y,eq:z} are given as
\begin{align}
    \begin{pmatrix}
           \delta x' \\
            \delta y' \\
           \delta z'
         \end{pmatrix} &= \mathcal{J}
         \begin{pmatrix}
           \delta x \\
            \delta y \\
           \delta z
         \end{pmatrix},
\end{align}
where $\mathcal{J}$ is the $3\times 3$ Jacobian matrix with components
\begin{subequations}
\begin{align}
&\mathcal{J}_{11}  = \dfrac{(3 x+2) \left[3 x z (2 \alpha +1) +2 y\right]+2 z-3}{2 (1-z)}\,, \\
&\mathcal{J}_{12}  = \dfrac{(x+1) (3 x+1)}{2 (1-z)} \,, \\
&\mathcal{J}_{13}  = \dfrac{(x+1) \left[3 x (2 \alpha  x+x+y)+y-1\right]}{2 (1-z)^2}\,, \\
&\mathcal{J}_{21}  = \dfrac{3}{2}\left[\dfrac{4 \alpha x z (y-2) +(y-2 z) (2 x z+y)}{1-z}\right], \\
&\mathcal{J}_{22}  = \dfrac{y (6 x-4 z+6)+3 x z (2 \alpha  x+x-2)-3}{2 (1-z)}\,,  \\
&\mathcal{J}_{23}  = \dfrac{1}{2(1-z)^2}\Big\{6 x^2 \left[(z-2) z-2 \alpha \right]+(3 x+1) y^2+2  \nonumber \\
&\hspace{3cm} +3 y \left[x (2 \alpha  x+x-2)-1\right]\Big\}, \\
&\mathcal{J}_{31}  = 6\alpha z\,, \quad \mathcal{J}_{32} = z\,, \quad \mathcal{J}_{33}= 6 \alpha  x+y\,. 
\end{align}
\end{subequations}
At the first critical point, the eigenvalues of $\mathcal{J}$ are
\begin{equation}
(\mu_1,  \mu_2,  \mu_3)_{I}=\left(-\frac{3}{2}, \, -\frac{3}{2}, \, -2 \alpha\right).
\end{equation}
If $\alpha>0$, the eigenvalues are all negative and, thus, the fixed point is stable and represents an attractor (see left panel of \Cref{fig:attractors}). If $\alpha<0$, the eigenvalues have mixed signs and the fixed point is a saddle point.

At the second and third critical points, the eigenvalues read
\begin{equation}
(\mu_1,  \mu_2,  \mu_3)_{II,III}=\left(0, \, 0, \, 3-3 \alpha \pm \sqrt{3\alpha  (3 \alpha -2)}\right),
\end{equation}
implying that they are unstable points.

At the fourth critical point, we find
\begin{equation}
(\mu_1,  \mu_2,  \mu_3)_{IV}=\left(\frac{2-3 \alpha }{1-3 \alpha}, \, \frac{2-3 \alpha }{1-3 \alpha }, \, \frac{5-9 \alpha }{1-3 \alpha }\right).
\end{equation}
If $1/3<\alpha <5/9$, the eigenvalues are all negative and the fixed point is stable, while for $5/9<\alpha <2/3$ we have a saddle point. Specifically, the de Sitter solution obtained for $\alpha=1/2$ implies
\begin{equation}
(\mu_1,  \mu_2,  \mu_3)_{IV}^{(dS)}=(-1,\,-1,\,-1)\,,
\end{equation}
suggesting an attractor point (see right panel of \Cref{fig:attractors}). 

The eigenvalues corresponding to the fifth critical point are not straightforward to calculate.
It can be shown that there is a saddle point for $\alpha<0\, \vee\, 0<\alpha<5/9\, \vee\, \alpha >3/4$. Instead, for the de Sitter solution obtained for $\alpha=3/4$, we find 
\begin{equation}
(\mu_1,  \mu_2,  \mu_3)_{V}^{(dS)}=(0,\,0,\,0)\,,
\end{equation}
showing that the point is unstable. 

\begin{figure}
\includegraphics[width=3.3in]{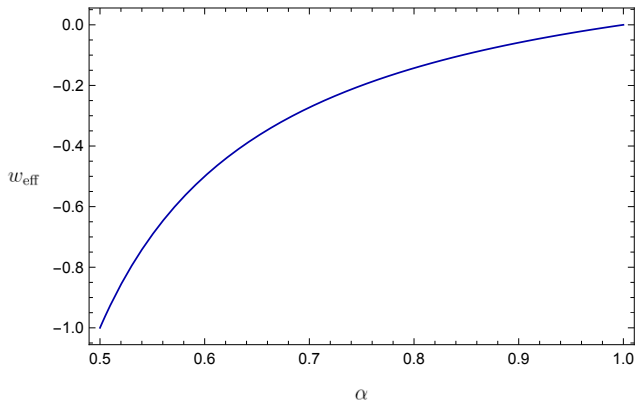}
\caption{Effective EoS parameter for the fourth critical point for the $f(\phi)=f_0e^{\alpha \phi}$ model (c.f. \Cref{eq:w_eff}).}
\label{fig:weff}
\end{figure}

It  is worth discussing  the outcomes of the phase-space analysis related to the  effective EoS  related to the  non-local model considered here. In particular, it is useful to infer theoretical bounds for the free parameter $\alpha$. 
To do so, let us  notice that no stable cosmological solutions may occur in the case of  a universe depleted of matter, with a pure cosmological constant as the only fluid contribution.
The physical solutions summarized in \Cref{tab:critical}, obtained by the autonomous dynamical system, refer to the critical points of the corresponding phase space, and do not provide further information on the present value of the matter density, $\Omega_m\simeq0.3$, as constrained from current observations.

It is important  to compare the non-local model with the $\Lambda$CDM predictions and with dark fluid scenarios. If, on the one hand, $w_\text{eff}=-1$ is the  case for the net equation of state of the $\Lambda$CDM model, on the other hand the dark fluid\footnote{The dark fluid scenario fully degenerates with the $\Lambda$CDM model, but the former contemplates a universe made of a single fluid, differently from the latter \cite{df3,df4}.}  can provide an EoS $w_\text{eff}\simeq -3/4$ \cite{df1,df2}. For comparison, let us consider the solutions we found in correspondence of the fourth critical point (see \Cref{tab:critical}). In this case, the condition $-1\leq w_\text{eff}<0$ places the constraint $1/2 \leq \alpha < 1$, as shown in \Cref{fig:weff}. However, as previously discussed, the stability of the fourth fixed point limits this range to $1/3 <\alpha < 5/9$, so that plausible values of $\alpha$ lie in the narrow interval $[0.5,\,0.55]$. Besides  such fine-tuning issue, we ought to bear in mind that, in this case, matter is identically zero, and thus the resulting scenario is not particularly suitable for cosmological purposes.

On the other hand, to find adequate values of $\alpha$, one would have to numerically solve the dynamical system from early epochs to current times. This may be realized only after deriving suitable initial conditions over the scalar fields.
As evidenced by the ghost-free conditions reported in the Appendix, we notice that all the values of $\alpha$ discussed in our analysis are admitted in principle, as long as one chooses the appropriate sign of the overall constant of the exponential coupling.

\section{Discussion and conclusions}
\label{sec:conclusions}

In this paper,  we analyzed the cosmological features of an effective gravitational action containing non-local terms as function of  the inverse d'Alembert operator. 
Specifically, we considered a Lagrangian density of the form $f(R,\Box^{-1}R)$, which can be thought as a straightforward non-local extension of $f(R)$ gravity \cite{Bajardi2}. Then, after reducing the action to a two-scalar tensor theory with a scalar field non-minimally coupled to gravity, we worked out the field equations and the equations governing the evolution of the scalar fields in  a homogeneous and isotropic universe.

To investigate the cosmic dynamics, we made use of dimensionless variables and recast the cosmological equations as an autonomous system of differential equations, which determine the behavior of the effective EoS. Then, we studied the solutions of the obtained system accounting for the non-local contributions through the exponential function $f(\phi)\propto e^{\alpha \phi}$, which naturally emerge from  symmetries of the gravitational Lagrangian \cite{Bajardi2}. We thus performed a phase-space analysis in terms of critical points and their stability. From the study of linear perturbations of the dynamical system and the sign of the eigenvalues of the Jacobian matrix evaluated at each fixed point, we searched for the existence of unstable, saddle points and attractor solutions at late-times, depending on value of the parameter $\alpha$. The matter-dominated universe and the accelerated universe solutions are particularly interesting. They represent cosmological attractors if $\alpha>0$ and $1/3<\alpha <5/9$, respectively. It is possible to show that, for $\alpha=1/2$, one finds a de Sitter universe  dominated by the cosmological constant, with the corresponding fixed point being a stable attractor. 
Hence, we discussed the physical predictions of the non-local scenario compared to the $\Lambda$CDM and dark fluid models.

In order to confirm the viability of the model considered here, it would be interesting to perform a direct comparison with the cosmological observations through a numerical analysis that includes Bayesian methods. This would provide constraints on the parameter $\alpha$ and fully determines the cosmic evolution. Furthermore, alternative functional forms for the non-local terms could be considered with the aim of checking for possible biases in the theoretical results. These questions will be the subjects for future investigations. As a final remark, it is worth noticing that the presence of non-local terms in the gravitational Lagrangian is a natural way to modulate the transition from matter-dominated to late accelerated expansion giving the possibility to fix cosmological tensions and other recent observational issues \cite{Lambiase}.

\begin{acknowledgements}
S.C. and R.D. acknowledge the support of  Istituto Nazionale di Fisica Nucleare (\textit{iniziative specifiche} QGSKY and MoonLight2). O.L. acknowledges funds from the Ministry of Education and Science of the Republic of Kazakhstan, Grant: IRN AP08052311.
\end{acknowledgements}

\appendix*
\section{Ghost-free conditions for the theory}
\label{appendix}

Here, we discuss the ghost-free conditions for the theory under study.
For this purpose, we shall consider a conformal transformation to the Einstein frame, namely $\tilde{g}_{\mu\nu}=\omega^2 g_{\mu\nu}$, so that
\begin{equation}
\tilde{R}=\frac{1}{\omega^2}\left[R-6\left(\Box\ln \omega+g^{\mu\nu}\nabla_\mu\ln\omega\nabla_\nu\ln\omega\right)\right],
\end{equation}
where the conformal factor is $\omega^{-2}=1+f(\phi)-\xi$.
Thus, in the Einstein frame, action \eqref{eq:action_localized} reads 
\begin{align}
&S =  \int d^4 x \sqrt{-g} \Big\{\dfrac{1}{2\kappa}\left[R-6g^{\mu\nu}\nabla_\mu\ln\omega\nabla_\nu\ln\omega\right. \nonumber \\
&\hspace{3cm}\left.-\omega^2g^{\mu\nu}\nabla_\mu\xi\nabla_\nu\phi \right]+\omega^4\mathcal{L}_m \Big\}.
\label{eq:action_conformal}
\end{align}
where we have neglected the term $\Box\ln \omega$ as it represents a total divergence.
It is then convenient to introduce a new field $\varphi$,
\begin{equation}
\varphi\equiv \ln\omega=-\frac{1}{2}\ln(1+f(\phi)-\xi)\,,
\end{equation}
which imposes the following condition:
\begin{equation}
1+f(\phi)-\xi>0 \Longleftrightarrow \psi>-1\,. 
\label{eq:first condition}
\end{equation}
One thus can write action \eqref{eq:action_conformal} as
\begin{align}
&S=\int d^4 x \sqrt{-g} \Big\{\dfrac{1}{2\kappa}\left[R-6g^{\mu\nu}\nabla_\mu\varphi\nabla_\nu\varphi-e^{2\varphi}g^{\mu\nu}\nabla_\mu\xi\nabla_\nu\phi\right] \nonumber \\
&\hspace{2.7cm}+e^{4\varphi}\mathcal{L}_m \Big\}.
\end{align}
Finally, using the relation $\xi=1+f(\phi)-e^{-2\varphi}$, we obtain 
\begin{align}
&\hspace{-0.3cm}S=\int d^4 x \sqrt{-g} \Big\{\dfrac{1}{2\kappa}\left[R-6\nabla^\mu\varphi\nabla_\mu\varphi-2\nabla^\mu\varphi\nabla_\mu\phi\right.  \nonumber \\
&\hspace{2.5cm}\left.-e^{2\varphi}f'(\phi)\nabla^\mu\phi\nabla_\mu\phi\right]+e^{4\varphi}\mathcal{L}_m \Big\}.
\end{align}

The condition to ensure the absence of a ghost is obtained by  requiring the positivity of the determinant of the kinetic term matrix, namely
\begin{equation}
\begin{vmatrix}
     6 & 1\\ 
     1 & e^{2\varphi}f'(\phi)\\
\end{vmatrix}
=6e^{2\varphi}f'(\phi)-1>0\,,
\end{equation}
which translates into 
\begin{equation}
f'(\phi)>\frac{e^{-2\varphi}}{6}=\frac{1+f(\phi)-\xi}{6}\,.
\end{equation}
Therefore, in virtue of \eqref{eq:first condition}, we find the  condition
\begin{equation}
f'(\phi)>0\,.
\end{equation}

For the specific case considered in the present study, namely Eq. \eqref{ansatz}, the above condition provides the constraints  $f_0>0$ ($f_0<0$) for $\alpha>0$ ($\alpha<0$). This result shows that it is possible to choose the free parameter $f_0$ such that 
both positive and negative values of $\alpha$ are admitted without introducing ghosts.

\end{document}